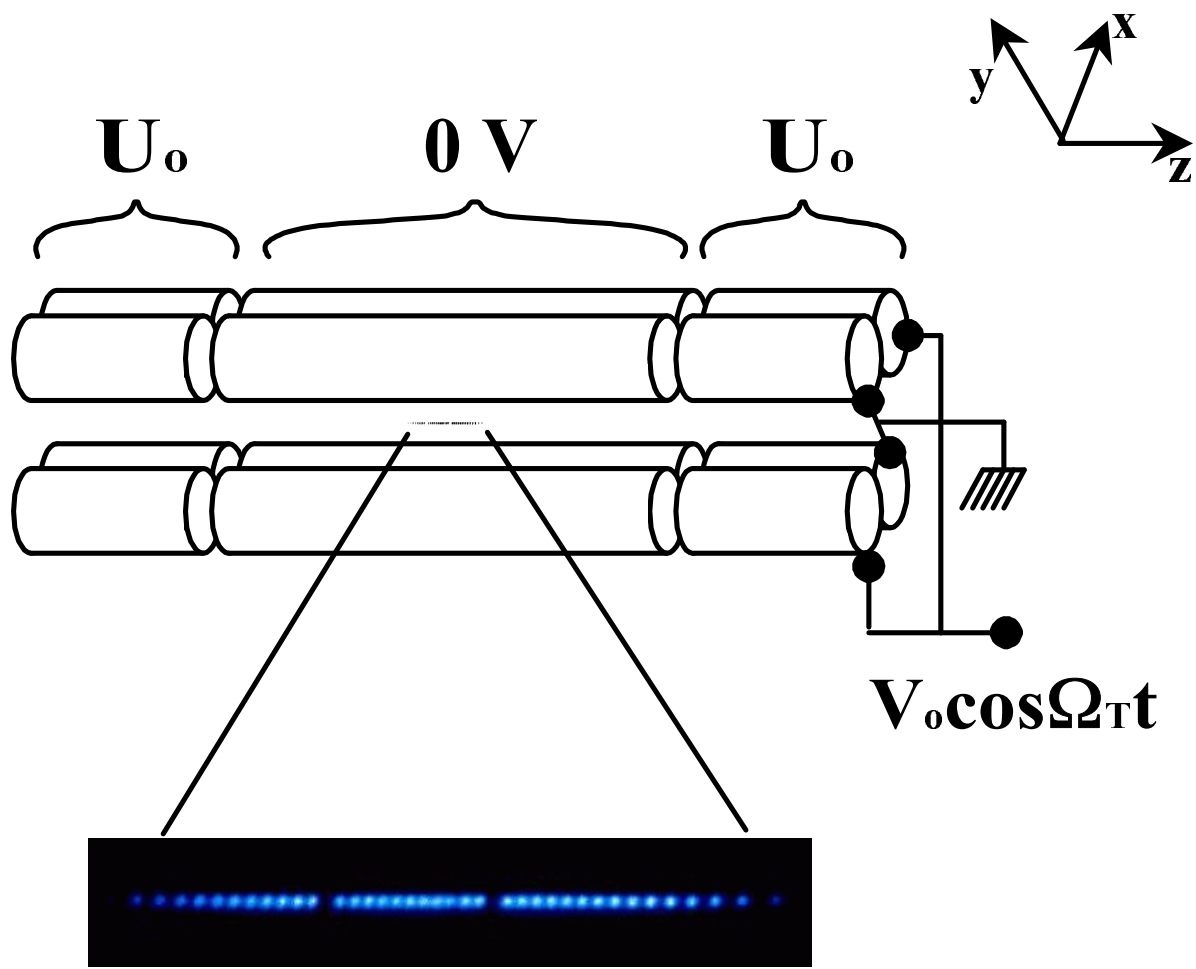

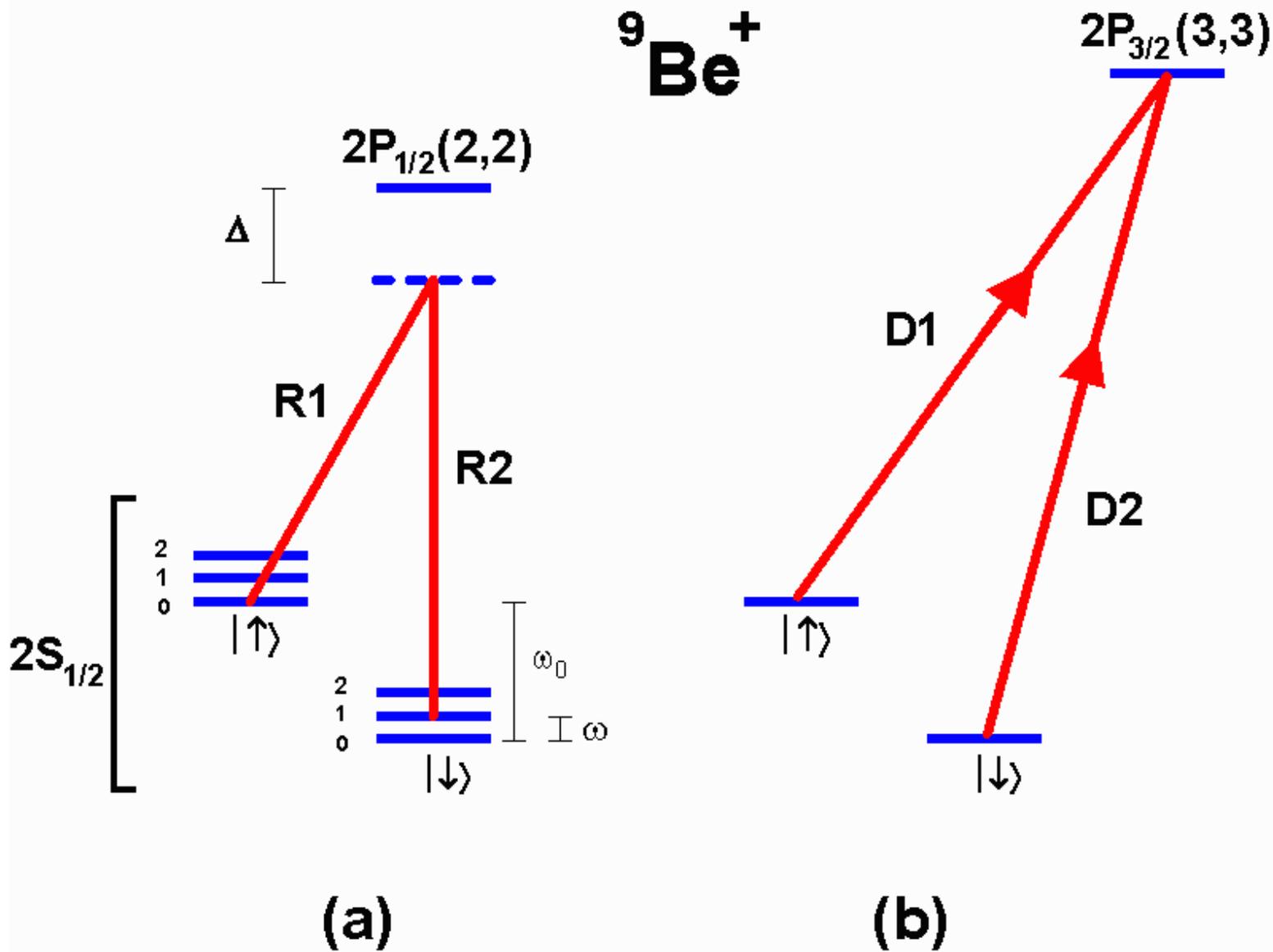

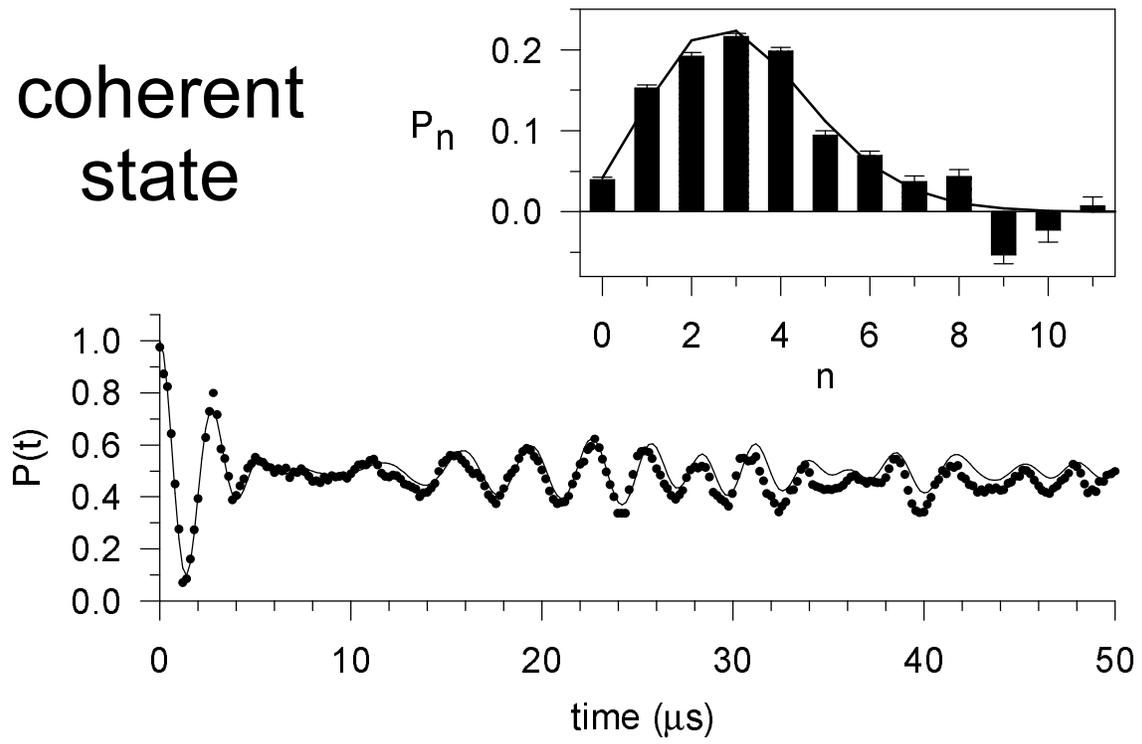

Fig. 3.

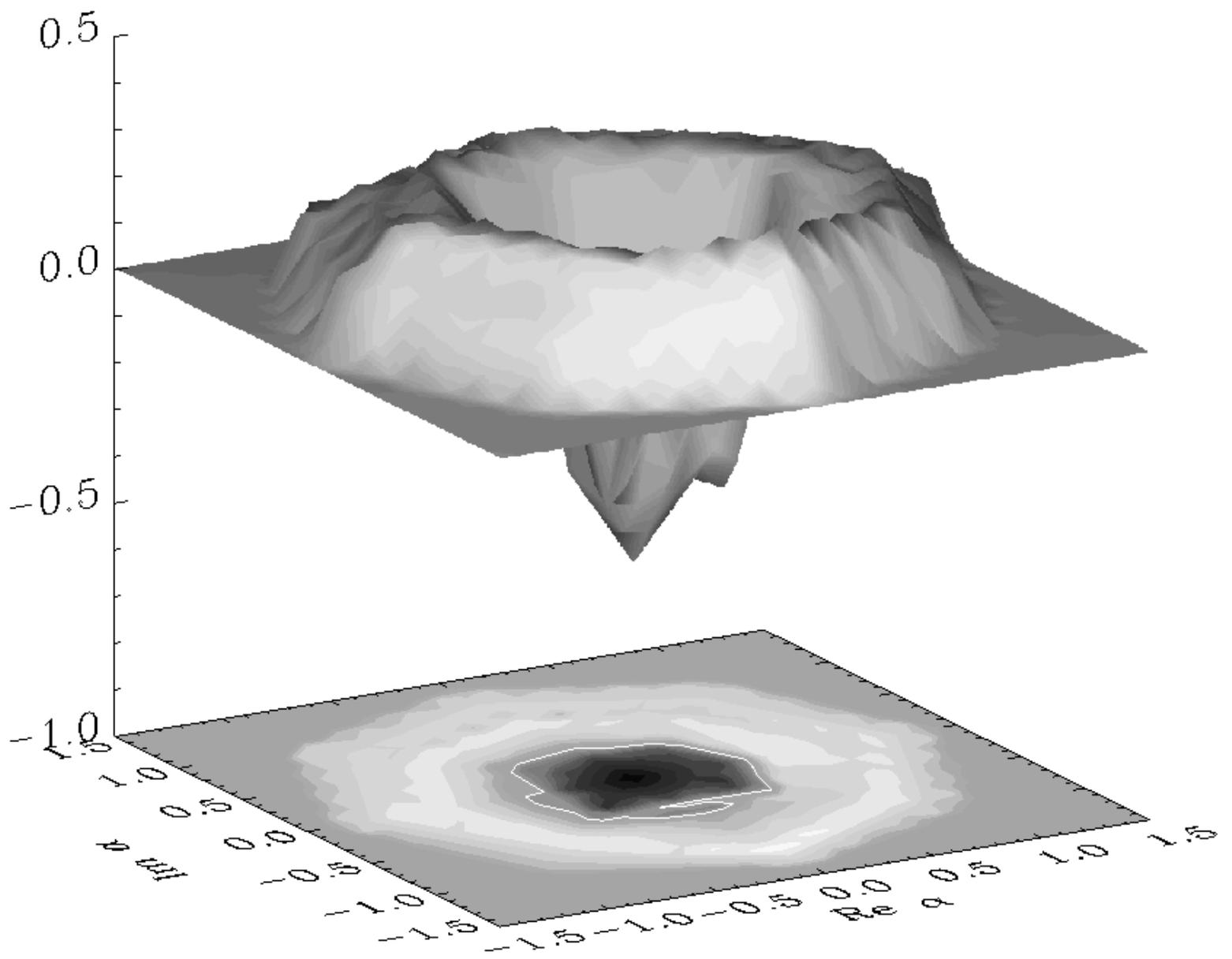

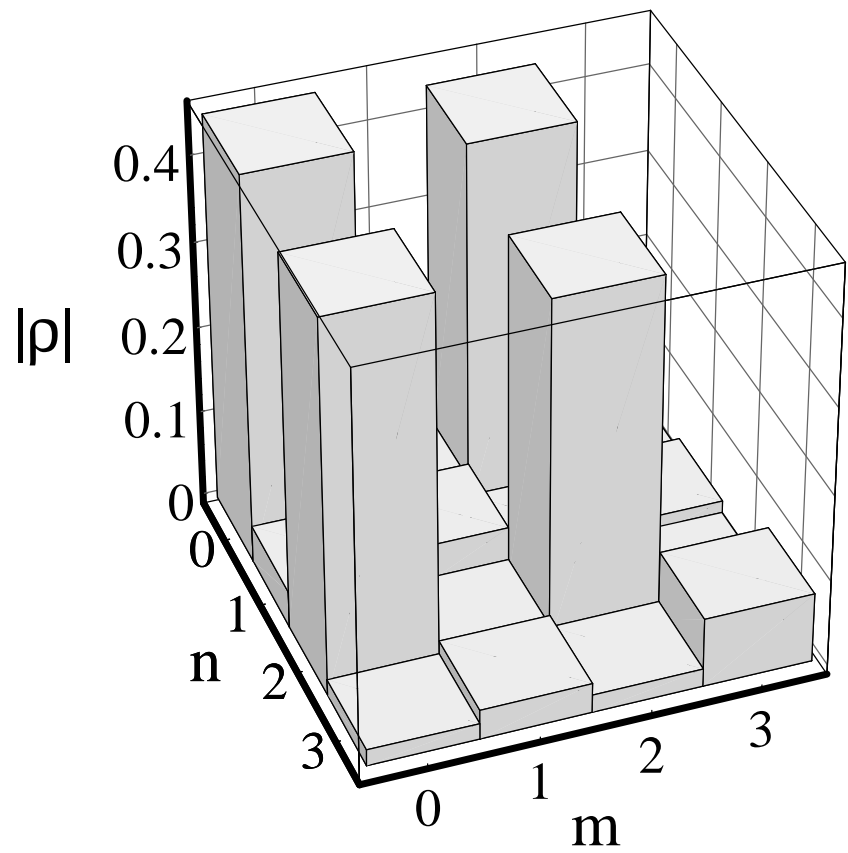

# Quantum State Manipulation
# Of Trapped Atomic Ions[†]

D.J. Wineland, C. Monroe, D.M. Meekhof,
B.E. King, D. Leibfried, W.M. Itano, J.C. Bergquist,
D. Berkeland, J.J. Bollinger, and J. Miller

Ion Storage Group, Time and Frequency Division
NIST, Boulder, CO, 80303, USA

Abstract:

A single laser-cooled and trapped $^9Be^+$ ion is used to investigate methods of coherent quantum-state synthesis and quantum logic. We create and characterize nonclassical states of motion including "Schrödinger-cat" states. A fundamental quantum logic gate is realized which uses two states of the quantized ion motion and two ion internal states as qubits. We explore some of the applications for, and problems in realizing, quantum computation based on multiple trapped ions.

# 1. INTRODUCTION

Currently, a major theme in atomic, molecular, and optical physics is coherent control of quantum states. This theme is manifested in a number of topics such as atom interferometry, Bose-Einstein condensation and the atom laser, cavity QED, quantum computation, quantum-state engineering, wavepacket dynamics, and coherent control of chemical reactions.

Here, we report related trapped-ion research at NIST which is devoted primarily to achieving quantum computation based on a scheme proposed by Cirac and Zoller [1]. As progress towards this goal, we report experiments on (1) the dynamics of a two-level atomic system coupled to harmonic atomic motion, (2) the creation and characterization of nonclassical states of motion such as "Schrödinger-cat" superposition states, and (3) simple quantum logic gates which are the precursors to Cirac and Zoller's ion-trap quantum computer.

# 2. TRAPPED ATOMIC IONS

Because of their overall electric charge, atomic or molecular ions can be confined by particular arrangements of electromagnetic fields for relatively long periods of times (hours or longer) with relatively small perturbations to their internal energy level structure. For studies of ions at low kinetic energy (< 1 eV), two types of trap are commonly used - the Penning and Paul traps. The Penning trap uses a combination of static electric and magnetic fields for confinement, and the Paul, or rf trap confines ions primarily through ponderomotive fields generated by inhomogeneous oscillating electric fields. The operation of these traps is discussed in various reviews (for example, see Refs. 2 - 4) and in a recent book [5]. For brevity, we discuss one trap configuration which is useful for the topics discussed in this paper.

In Fig. 1 we show schematically a "linear" Paul trap. This trap is based on the one described in Ref. [6], which is derived from the original design of Paul [7]. It is basically a quadrupole mass filter which is plugged at the ends with static electric potentials. An oscillating potential $V_o \cos(\Omega_T t)$ is applied between diagonally opposite rods, which are fixed in a quadrupolar configuration, as indicated in Fig. 1. We assume that the rod segments along the z (horizontal) direction are coupled together with capacitors (not shown), so the rf potential is constant as a function of z. Near the axis of the trap this creates an oscillating potential of the form

$$V \simeq \frac{V_o}{2}\cos(\Omega_T t)\left(1 + \frac{x^2 - y^2}{R^2}\right), \qquad (1)$$

where R is equal to the distance from the axis to the surface of the electrode [8]. This gives rise to a (harmonic) ponderomotive potential in the x-y direction which, for a single ion (or the center-of-mass (COM) motion of a group of ions), yields the motion

$$x(t) \simeq X(1 + (q_x/2)\cos\Omega t)\cos(\omega_x t + \phi_x),$$
$$y(t) \simeq Y(1 + (q_y/2)\cos\Omega t)\cos(\omega_y t + \phi_y), \qquad (2)$$

where $q_x = -q_y \equiv 2qV_o/(mR^2\Omega_T^2) = 2\sqrt{2}\omega_x/\Omega_T$, $\omega_x = \omega_y = \omega = qV_o/(\sqrt{2}mR^2\Omega_T)$, and m and q are the ion mass and charge. Equations (2) are valid for the typical conditions that $|q_x|, |q_y| \ll 1$. In this approximation, if we neglect the micromotion (the $(q_i/2)\cos\Omega t$ terms in Eq. (2)), the ion behaves as if it were confined in a pseudopotential $\Phi_p$ given by

$$q\Phi_p = \frac{1}{2}m\omega_i^2(x^2 + y^2), \qquad (3)$$

in which the ion oscillates with secular frequency $\omega_i$. To provide confinement along the z direction, static potentials are applied to the end segments of the rods relative to the central segments. Near the center of the trap, this provides a static harmonic well along z. This necessarily weakens the well in the radial direction [6], but we will assume that the binding in the radial direction is large compared to that along z, so this effect is small. Figure 1 also shows an image of a "string" of $^{199}$Hg$^+$ ions which are confined near the z axis of the trap described in Ref. 9. This string of ions might be used as a register in a quantum computer [1].

When this kind of trap is installed in a high-vacuum apparatus (pressure < $10^{-8}$ Pa), ions can be confined for days with minimal perturbations to their internal structure. Collisions with background gas are infrequent (the mean time between collisions is typically more than a few minutes). Even though the ions interact strongly through their mutual Coulomb interaction, the fact that the ions are localized necessarily means that $\langle \vec{E} \rangle = 0$ at the positions of the ions; therefore electric field perturbations occur only in second order. These perturbations are typically very small [10]. Magnetic field perturbations are important; however, if we operate at fields where the energy separation between levels is at an extremum with respect to field, the coherence time for superposition states of the two levels can be very long. For example, in a $^9$Be$^+$ (Penning trap) experiment operating in a field of 0.82 T, the coherence time of a superposition between hyperfine states exceeded 10 minutes [11]. As described below, we will be interested in exciting coherently the quantized modes of the ion motion in the trap. Not surprisingly, the coupling to the environment is relatively strong because of the ions' charge. One measure of the decoherence rate due to environmental coupling is the rate of transitions between the ions' quantized oscillator levels. Transition times from the zero-point energy level (achieved from laser cooling) to the first excited motional state have been measured for single $^{198}$Hg$^+$ ions to be about 0.15 s [12] and for single $^9$Be$^+$ ions to be about 1 ms [13].

Ion trap experiments can achieve high detection efficiency of the ion's internal states. Unit detection efficiency has been achieved in previous experiments on "quantum jumps" [14] where the internal state of the ion is indicated by light scattering (or lack thereof), correlated with the ion's internal state.

In the experiments we discuss here, a single $^9$Be$^+$ ion was stored in a conventional Paul trap [2-4] where the ion's confinement is characterized by secular oscillation frequencies ($\omega_x$, $\omega_y$, $\omega_z$)/$2\pi \approx$ (11, 19, 29) MHz along the three principal axes of the trap [15]. The future goal of the work will be able to extend these experiments to the case of multiple ions in a linear trap (Fig. 1).

## 3. ENTANGLED QUANTUM STATES

An entangled quantum state is one where the wave function of the overall system cannot be written as a product of the wave functions of the subsystems. In this case, a measurement on one of the subsystems will affect the state of the other subsystems. For example, consider a two-level atom bound in a 1-D harmonic well. Suppose we can create the state

$$\Psi = \frac{1}{\sqrt{2}}(|\downarrow\rangle|n\rangle + e^{i\phi}|\uparrow\rangle|n'\rangle), \qquad (4)$$

where the kets $|\downarrow\rangle$ and $|\uparrow\rangle$ denote the two internal eigenstates of the atom (here, we use the spin-½ analog to a two-level system: $\sigma_z|\uparrow\rangle = +|\uparrow\rangle$, etc. ), the second ket denotes a harmonic oscillator eigenstate $|n\rangle$, and $\phi$ is a (controlled) phase factor. This state is entangled because, if we measure the motional eigenstate of the atom and find it to be in level n, then it must also be found in the $\downarrow$ internal state if we measure $\sigma_z$. Similarly, if we find the atom in the n' motional state, it must be found in the $\uparrow$ internal state. Such correlations are central to "EPR" experiments [16]. Another interesting state is the state for N two-level atoms

$$\Psi = \frac{1}{\sqrt{2}}(|\downarrow\rangle_1|\downarrow\rangle_2...|\downarrow\rangle_N + e^{i\phi}|\uparrow\rangle_1|\uparrow\rangle_2...|\uparrow\rangle_N), \qquad (5)$$

where the subscript i (= 1, 2, ..., N) denotes the ith atom. This state is "maximally entangled" in the sense that a measurement of $\sigma_z$ on any atom automatically determines the value of $\sigma_z$ of all other atoms.

## 4. LASER COOLING TO THE ZERO-POINT STATE

As a starting point for the experiments discussed in this paper, we want to prepare the $^9$Be$^+$ ion in a particular internal state and in the lowest quantized level of vibrational motion in the trap - the zero-point state.

The ion can be optically pumped into a particular internal state using polarized light. Preparation in the zero-point state of motion is achieved with laser cooling. We cool a single, trapped $^9$Be$^+$ ion to near the zero-point energy using resolved-sideband laser cooling [17] which can be explained as follows: Consider a two-level atom characterized by a resonant (optical) transition frequency $\omega_0$ and radiative linewidth $\gamma$. For simplicity, we assume the atom is confined by a 1-D harmonic well of vibration frequency $\omega_x \gg \gamma$. If a laser beam (frequency $\omega$) is incident along the direction of the atomic motion, the bound atom's absorption spectrum is composed of a "carrier" at frequency $\omega_0$ and resolved (since $\omega_x \gg \gamma$) frequency-modulation sidebands that are spaced by $\omega_x$ and are generated from the Doppler effect. Cooling occurs if the laser is tuned to a lower sideband, for example, at $\omega = \omega_0 - \omega_x$. In this case, photons of energy $\hbar(\omega_0 - \omega_x)$ are absorbed, and spontaneously emitted photons of average energy $\hbar\omega_0$ return the atom to its initial internal state (assuming $\hbar\omega_x$ is much greater than the photon recoil energy for the bound atom). This reduces the atom's kinetic energy by $\hbar\omega_x$ per scattering event. Cooling proceeds until the atom's mean vibrational quantum number in the harmonic well is given by $\langle n \rangle_{min} \simeq (\gamma/2\omega_x)^2 \ll 1$. Experimentally, we find it convenient to use two-photon stimulated Raman transitions for cooling (as described below), but the basic idea for cooling is essentially the same as for single-photon transitions described here.

## 5. COUPLING BETWEEN INTERNAL AND MOTIONAL STATES

To achieve both laser cooling and entanglement, we need to provide a coupling between internal and motional quantum states. This can be achieved with the application of inhomogeneous (classical) electromagnetic fields. For example, consider an atom confined in a 1-D harmonic potential. The atom's dipole moment $\boldsymbol{\mu}$ is assumed to couple to an electric field $E(x,t)$ through the Hamiltonian

$$H_I = -\boldsymbol{\mu} E(\boldsymbol{x},t) = -\boldsymbol{\mu}\left[E(x=0,t) + \frac{\partial E}{\partial x}\boldsymbol{x} + \frac{1}{2}\frac{\partial^2 E}{\partial x^2}\boldsymbol{x^2} + \cdots\right]. \quad (6)$$

We have $\boldsymbol{\mu} \propto \sigma_+ + \sigma_-$, where $\sigma_+$ and $\sigma_-$ are the raising and lowering operators for the internal levels (in the spin-½ analog). In Eq. (6), the position $\mathbf{x}$ is an operator representing the position of the atom, which we write as $\mathbf{x} = x_o(\mathbf{a} + \mathbf{a}^\dagger)$, where $\mathbf{a}$ and $\mathbf{a}^\dagger$ are the usual harmonic oscillator lowering and raising operators, and $x_o$ is the rms spread of the n=0 zero-point state of motion.

As a simple example, suppose the field is static and the motional oscillation frequency $\omega_x$ of the atom is equal to the resonance frequency $\omega_o$ of the internal state transition. In its reference frame, the atom experiences an oscillating field due to the motion through the inhomogeneous field. Since $\omega_x = \omega_o$, this field resonantly drives transitions between the internal states. If the extent of the atom's motion is small enough that we need only consider the first two terms in Eq. (6), the second term of $H_I$ can be approximated as $\hbar\Omega(\sigma_+\mathbf{a} + \sigma_-\mathbf{a}^\dagger)$ (in the interaction frame and using the rotating wave approximation), where $\Omega$ is a proportionality constant containing the field gradient $\partial E/\partial x$. This Hamiltonian is equivalent to the Jaynes-Cummings Hamiltonian of cavity-QED [18] which describes the coupling between a two-level atom and a single mode of the radiation field. This analogy has been pointed out in various papers [19-22]; for a review, see Ref. 23 and further references in Ref. 22.

*5.1    Realization of a Jaynes-Cummings-type coupling for a trapped $^9Be^+$ ion*

The relevant energy-level structure of $^9Be^+$ is summarized in Fig. 2. Because the ion is trapped, the internal $^9Be^+$ electronic states must include the ladder of external harmonic oscillator levels of energy $E_n = \hbar\omega(n+\tfrac{1}{2})$, where we have considered only the x-dimension of the oscillator ($\omega \equiv \omega_x$) and its associated quantum number $n \equiv n_x \in (0, 1, 2, ...)$. The two internal levels of interest are the $^2S_{1/2}$ ground state hyperfine levels $|F=2, m_F=2\rangle$ (denoted by $|\downarrow\rangle$) and $|F=1, m_F=1\rangle$ (denoted by $|\uparrow\rangle$), which are separated in frequency by $\omega_o/2\pi \approx 1.25$ GHz. The other Zeeman levels (not shown) are resolved from the $|\downarrow\rangle$ and $|\uparrow\rangle$ states by the application of a $\approx 0.2$ mT magnetic field [13,22].

Strong field gradients are obtained with laser fields from the $e^{ikx}$ factor in $E_{laser} \propto e^{i(kx - \omega t)}$. In our experiment, the field corresponding to that in Eq. (6) is provided by two laser fields (R1 and R2 of Fig. 2a) which drive stimulated-Raman transitions between the levels of interest. The use of stimulated-Raman transitions has some technical advantages, but is formally equivalent to driving a narrow single-photon transition. Raman transitions between the $|\downarrow\rangle$ and $|\uparrow\rangle$ states can be driven by tuning the difference frequency of R1 and R2 to be near $\omega_o$. The two Raman beams ($\lambda \approx 313$ nm) are generated from a single laser source and an acousto-optic modulator, allowing excellent stability of their relative frequency and phase. Both beams are detuned $\Delta/2\pi \approx -12$ GHz from the excited $^2P_{1/2}$ electronic state (radiative linewidth $\gamma/2\pi \approx 19.4$ MHz), and the polarizations are set to couple through the $^2P_{1/2}(F=2, m_F=2)$ level (the next nearest levels are the $^2P_{3/2}$ states which are over 200 GHz away and can be neglected). Because $\Delta \gg \gamma$, the excited $^2P$ state can be adiabatically eliminated in a theoretical description, resulting in a coupling between the two ground states which exhibits a linewidth inversely proportional to the interaction time. When R1 and R2 are applied to the ion with wavevector difference $\delta\vec{k} = \vec{k}_1 - \vec{k}_2$ along the x direction, the effective coupling Hamiltonian in the rotating-wave approximation is given by

$$H_I = g\left(\sigma_+ e^{i\eta(\mathbf{a}^\dagger + \mathbf{a}) - i\delta t} + \sigma_- e^{-i\eta(\mathbf{a}^\dagger + \mathbf{a}) + i\delta t}\right). \qquad (7)$$

The coupling strength g depends on $\Delta$ and the intensity of the laser beams, $\eta = |\delta\vec{K}|x_0 \simeq 0.2$ is the Lamb-Dicke parameter, $x_0 = (\hbar/2m\omega_x)^{1/2} \approx 7$ nm, and $\delta$ is the difference between the relative frequency of the two Raman beams and $\omega_o$. Setting $\delta\vec{K}$ to be parallel to the x axis of the trap yields almost no coupling between the internal states and motion in the y and z directions.

If $\delta = \omega_x(n'-n)$, transitions are resonantly driven between the levels $|\downarrow,n\rangle$ and $|\uparrow,n'\rangle$ at a rate $\Omega_{n,n'} \propto g$ which is dependent on n and n' [17,22,24] (analogous to Franck-Condon factors in molecular spectroscopy). Starting from the $|\downarrow\rangle|n\rangle$ state, application of a "Rabi $\pi$" pulse coherently transfers the ion to the $|\uparrow\rangle|n'\rangle$ state; this corresponds to applying the Raman beams for a duration $\tau$ such that $2\Omega_{n,n'}\tau = \pi$. If we apply the Raman beams for half of this time, we create the entangled state of Eq. (4). Here, we will assume the ion is confined in the Lamb-Dicke limit ($|\delta\vec{K}|\langle x^2\rangle^{1/2} << 1$) and will consider three transitions. The carrier, at $\delta = 0$, drives transitions between states $|\downarrow,n\rangle \leftrightarrow |\uparrow,n\rangle$ with Rabi frequency $\Omega_{n,n} = g$. The first red sideband, corresponding to $\delta = -\omega_x$, drives transitions between states $|\downarrow,n\rangle \leftrightarrow |\uparrow,n-1\rangle$ with Rabi frequency $\Omega_{n,n-1} = g\eta\sqrt{n}$. This coupling is analogous to the Jaynes-Cummings coupling in cavity QED [18] where energy is coherently exchanged between the internal state and single-photon radiation field. The first blue sideband, at $\delta = +\omega_x$, drives transitions between states $|\downarrow,n\rangle \leftrightarrow |\uparrow,n+1\rangle$ with Rabi frequency $\Omega_{n,n+1} = g\eta(n+1)^{1/2}$.

Laser cooling to the $|\downarrow\rangle|n=0\rangle$ state is accomplished in two stages. We first use first Doppler cooling [17] to cool the ion to a limit given by $\langle n\rangle \simeq 1$ for our experimental conditions. We then apply sideband laser cooling using stimulated Raman transitions [13] to achieve $\langle n\rangle \simeq 0$. For sideband laser cooling, Rabi $\pi$ pulses on the first red sideband ($|\downarrow\rangle|n\rangle \rightarrow |\uparrow\rangle|n-1\rangle$) are alternated with repumping cycles using nearly resonant radiation (beam D1 of Fig. 2b) - which results (most probably) in transitions $|\uparrow\rangle|n-1\rangle \rightarrow |\downarrow\rangle|n-1\rangle$. These steps are repeated (typically 5 times) until the ion resides in the $|\downarrow\rangle|0\rangle$ state with high probability ($> 0.9$).

As described below, from the $|\downarrow\rangle|0\rangle$ state we are able to coherently create states of the form $|\downarrow\rangle\Psi(x)$, where the motional state $\Psi(x) = \Sigma_n C_n \exp(-in\omega_x t)|n\rangle$ and the $C_n$ are complex. One way we can analyze the motional state created is as follows [22]: The Raman beams are applied to the ion for a time $\tau$, and the probability $P_\downarrow(\tau)$ that the ion is in the $|\downarrow\rangle$ internal state is measured. The experiment is repeated for a range of $\tau$ values. When the Raman beams are tuned to the first blue sideband, the expected signal is [22]

$$P_\downarrow(\tau) = \frac{1}{2}\left[1 + \sum_{n=0}^{\infty} P_n \cos(2\Omega_{n,n+1}\tau)e^{-\gamma_n \tau}\right], \qquad (8)$$

where $P_n \equiv |C_n|^2$ is the probability of finding the ion in state n and $\gamma_n$ are experimentally determined decay constants. The internal state $|\downarrow\rangle$ is detected by applying nearly resonant $\sigma^+$-polarized laser radiation (beam D2, Fig. 2b) between the $|\downarrow\rangle$ and $^2P_{3/2}(F=3, m_F=3)$ energy levels. Because this is a cycling transition, detection efficiency is near 1 [13,14,22]. The measured signal $P_\downarrow(\tau)$ can be inverted (Fourier cosine transform), allowing the extraction of the probability distribution of vibrational state occupation $P_n$. This signal does not reveal the phase coherences between the motional amplitudes; these must be verified separately [22,25]. The most complete characterization is achieved with a state reconstruction technique [26] discussed below.

## 6. CREATION OF COHERENT AND SCHRÖDINGER-CAT STATES OF MOTION

We have created and analyzed thermal, Fock, squeezed, coherent, Schrödinger-cat states, and other superpositions of Fock states [22,25,26]; here we briefly describe the creation and measurement of coherent and Schrödinger-cat states [22,25]. We note that a scheme recently proposed for producing arbitrary states of the electromagnetic field [27] should be directly applicable to the ion case for producing arbitrary states of motion.

A coherent state of motion is given by a particular superposition of motional eigenstates

$$\Psi(x) = |\alpha\rangle \equiv \exp(-|\alpha|^2/2) \sum_{n=0}^{\infty} \frac{\alpha^n}{\sqrt{n!}} |n\rangle, \qquad (9)$$

and corresponds to a displaced zero-point wave packet oscillating in the potential well with amplitude $2|\alpha|x_0$. From Eq. (8), $P_\downarrow(\tau)$ for a coherent state will undergo quantum collapses and revivals [28,29]. These revivals are a purely quantum effect due to the discrete energy levels and the narrow distribution of states [28,29].

We have produced coherent states of ion motion from the $|\downarrow\rangle|0\rangle$ state by applying a resonant (frequency $\omega_x$) force in the form of a uniform electric field acting on the ion charge, or in the form of a "moving standing wave" of laser radiation acting through the optical dipole force [22,25]. In Fig. 3, we show a measurement of $P_\downarrow(\tau)$ after creation of a coherent state of motion, exhibiting the expected collapse and revival signature. (For comparison, see the cavity-QED experiment of Ref. 30.) Eq. (8) is fitted to these data assuming a Poissonian distribution, allowing only $\langle n \rangle$ to vary. The inset shows the results of a separate analysis, which yield the probabilities of the Fock-state components, extracted by applying a Fourier cosine transform to $P_\downarrow(\tau)$ at the known frequencies as described above. These probabilities display the expected Poissonian dependence on n (Eq. (9)).

A Schrödinger-cat state is taken to be a coherent superposition of classical-like motional states. In Schrödinger's original thought experiment [31], he described how we could, in principle, entangle an internal state superposition of an atom with a macroscopic-scale superposition of a live and dead cat. In our experiment [25], we construct an analogous state, on a smaller scale, with a single atom. We create the state

$$\Psi = \frac{1}{\sqrt{2}}(|\downarrow\rangle|\alpha_1\rangle + e^{i\phi}|\uparrow\rangle|\alpha_2\rangle), \tag{10}$$

where $|\alpha_1\rangle$ and $|\alpha_2\rangle$ are coherent motional states (Eq. (9)) and $\phi$ is a (controlled) phase factor. The coherent states of the superposition are spatially separated by distances much greater than the size of the atom wavepacket which has an r.m.s. spread equal to $x_o$.

Analysis of this state is interesting from the point of view of the quantum measurement problem, an issue that has been debated since the inception of quantum theory by Einstein, Bohr, and others, and continues today [32]. One practical approach toward resolving this controversy is the introduction of quantum decoherence, or the environmentally induced reduction of quantum superpositions into classical statistical mixtures [33]. Decoherence provides a way to quantify the elusive boundary between classical and quantum worlds, and almost always precludes the existence of macroscopic Schrödinger-cat states, except for extremely short times. On the other hand, the creation of mesoscopic Schrödinger-cat states like that of Eq. (10) may allow controlled studies of quantum decoherence and the quantum-classical boundary. This problem is directly relevant to quantum computation, as we discuss below.

In our experiment, we create a Schrödinger-cat state of the single-ion $^9Be^+$ harmonic oscillator (Eq. (10)) with a sequence of laser pulses [25]. First, we create a state of the form $(|\downarrow\rangle + e^{i\xi}|\uparrow\rangle)|n=0\rangle/\sqrt{2}$ with a $\pi/2$ pulse on the Raman carrier transition (Sec. 5.1). To spatially separate the $|\downarrow\rangle$ and $|\uparrow\rangle$ components of the wave function, we apply a coherent excitation with an optical dipole force which, because of the polarization of the beams used to create the force, selectively excites the motion of only the $|\uparrow\rangle$ state. We then swap the $|\downarrow\rangle$ and $|\uparrow\rangle$ states with a carrier $\pi$ pulse and reapply the dipole force with a different phase to create the state of Eq. (10). In principle, if we could make $|\alpha_{1,2}|$ large enough, we could design a detector which could directly detect the (distinguishable) position of the particle and correlate it with a spin measurement [34]. Instead, to analyze this state in our experiment, we apply an additional laser pulse to couple the internal states, and we measure the resulting interference of the distinct wavepackets. With this interferometer, we can establish the correlations inherent in Eq. (10), the separation of the wavepackets, <u>and</u> the phase coherence $\phi$ between components of the wavefunction. These experiments are described in Ref. 25. The interference signal should be very sensitive to decoherence. As the separation $|\alpha_1 - \alpha_2|$ is made larger, decoherence is expected to exponentially degrade the fringe contrast if decoherence is due to radiative decay [33,35]. This type of decoherence has recently been observed for Schrödinger cats of the electromagnetic field [36].

Other experiments generate Schrödinger cats in the same sense as in our experiment. Examples are atom interferometers [37] and superpositions of electron wavepackets in atoms [38] (also, see additional citations in Ref. 25). However, as opposed to these experiments, the harmonic oscillator cat states of Eq. (10) do not disperse in time. This lack of dispersion provides a simple visualization of the "cat." For example, the ion in the trap can be likened to a marble rolling back and forth in a bowl. At certain times, the marble can be simultaneously on opposite sides of the bowl. These states should allow controlled studies of decoherence models and the tailoring of reservoirs which cause decoherence in the ion trap system [39].

## 7. COMPLETE RECONSTRUCTION OF THE MOTIONAL QUANTUM STATE

The controlled interaction of light and rf-electromagnetic fields with the trapped ion not only allows us to prepare very general states of motion, but also allows us to determine these quantum mechanical states. Few experiments have succeeded in determining the density matrices or Wigner functions of quantum systems [40,41]. Here we present the theory and experimental demonstration of two distinct schemes to reconstruct both the density matrix in the number state basis and the Wigner function of the motional state of a single trapped atom. Other proposed methods are discussed in Refs. and 42-45.

Our measurement techniques rely on our ability to coherently displace an "unknown" input state to several different locations in phase space. The technical realization of the displacements is described in the previous section in the context of the production of coherent states. After the coherent displacement we apply radiation on the first blue sideband for a time $\tau$, which induces a resonant exchange between the motional and internal degrees of freedom. For each coherent displacement, characterized by a complex number $\alpha$, a time series of measurements of the population $P_\downarrow(\tau)$ is made by monitoring the fluorescence produced in driving the resonant dipole cycling transition just as in Eq. (8). The internal state at $\tau=0$ is always prepared to be $|\downarrow\rangle$ for the various input states, so the signal averaged over many measurements is

$$P_\downarrow(\tau,\alpha) = \frac{1}{2}\left(1 + \sum_{n=0}^{\infty} Q_n(\alpha)\cos(2\Omega_{n,n+1}\tau)e^{-\gamma_n \tau}\right). \quad (11)$$

where $Q_n(\alpha)$ are the population probabilities of the displaced motional state. Without the coherent displacement we would just produce the signal of Eq. (8) and would get the populations of the motional eigenstates only. However, since we repeat these measurements for several magnitudes and phases of the coherently displaced state, we are able to extract information about the off-diagonal elements of the density matrix and can also reconstruct the Wigner function [46] from the measured displaced populations $Q_n(\alpha)$. These can be found by a Fourier cosine transform and are linearly related to the elements of the motional density matrix in the number state basis (a detailed description of these relations is given in Ref. 26). These relations can be numerically inverted to reconstruct the density matrix of the motional state in the number state basis using a general linear least-squares method.

As pointed out by several authors, quasiprobability distribution functions $F(\alpha,s)$ [46,47] such as the Wigner function can be characterized by a position in phase space $\alpha$ and a parameter s. These functions have a particularly simple representation when expressed in populations of coherently displaced number states $Q_k(\alpha)$ [47]:

$$F(\alpha,s) = \frac{1}{\pi} \sum_{n=0}^{\infty} [(s+1)/2]^n \sum_{k=0}^{n} (-1)^k \binom{n}{k} Q_k(\alpha). \qquad (12)$$

For $s = -1$, the sum reduces to one term, and $F(\alpha,-1) = Q_o(\alpha)/\pi$ gives the value of the Q-quasi-probability distribution at the complex coordinate $\alpha$ [43]. The Wigner function is given by $s = 0$. $F(\alpha,0) = W(\alpha)$ for every point $\alpha$ in the complex plane can be determined by the sum

$$W(\alpha) = \frac{2}{\pi} \left[ \sum_{n=0}^{\infty} (-1)^n Q_n(\alpha) \right]. \qquad (13)$$

In our reconstruction, the sum is carried out only to a finite $n_{max}$, depending on the estimated magnitude of the states to reconstruct [26].

In Fig. 4, we show the surface and contour plots of the Wigner function of an approximate $|n=1\rangle$ number state. The plotted points are the result of fitting a linear interpolation between the actual data points to a 0.1-by-0.1 grid. The octagonal shape is an artifact of the eight measured phases per radius. The white contour represents $W(\alpha)=0$. The negative values around the origin highlight the nonclassical character of this state. More recently, the negative Wigner function of a free atom has been observed [41].

As an example of a reconstructed number state density matrix, we show in Fig. 5 our result for a coherent superposition of $|n=0\rangle$ and $|n=2\rangle$ number states. This state is ideally suited to demonstrate the sensitivity of the reconstruction to coherences. Our result indicates that the prepared motional states in our system are very close to pure states.

## 8. QUANTUM LOGIC

Quantum computation has received a great deal of attention recently because of the algorithm proposed by Peter Shor for efficient factorization [48,49]. Accomplishing quantum factorization will be formidable with any technology; however, other applications of quantum logic may be more tractable. One possibility is described in Sec. 9. Ignacio Cirac and Peter Zoller [1] proposed an attractive scheme for a quantum computer which would use a string of ions in a linear trap as "qubits," similar to what is shown in Fig. 1. This proposal has stimulated experimental efforts in several laboratories, including those at Innsbruck University, Los Alamos National Laboratory, IBM (Almaden), Max-Planck Institute (Garching), NIST, and Oxford University.

Each qubit in a quantum computer could be implemented by a two-level atomic system; for the ith qubit, we label these states $|\downarrow\rangle_i$ and $|\uparrow\rangle_i$ as above. In general, any quantum computation can be comprised of a series of single-bit rotations and two-bit controlled-NOT (CN) logic operations [49-52]. We are interested in implementing these two operations in a system of $^9Be^+$ ions. Single-bit rotations are straightforward and correspond to driving Raman carrier transitions (Sec. 5.1) for a controlled time. Such rotations have been achieved in many previous experiments. In the following, we describe the demonstration of a CN logic gate with a single $^9Be^+$ ion [53].

*8.1     "Conditional dynamics" and a single-ion controlled-not logic gate*

The key to making a quantum logic gate is to provide conditional dynamics; that is, we desire to perform on one physical subsystem a unitary transformation which is conditioned upon the quantum state of another subsystem [49]. In the context of cavity QED, the required conditional dynamics at the quantum level has been demonstrated [54,55]. In NMR, conditional dynamics and CN operations have been realized [56,57]. For trapped ions, conditional dynamics at the quantum level has been demonstrated in verifications of zero-point laser cooling where absorption on the red sideband depended on the motional quantum state of the ion [12,13]. Recently, we have demonstrated a CN logic gate; in this experiment, we also had the ability to prepare arbitrary input states to the gate (the "keyboard" operation of step (2a) below).

A two-bit quantum CN operation provides the transformation:

$$|\epsilon_1\rangle|\epsilon_2\rangle \rightarrow |\epsilon_1\rangle|\epsilon_1 \oplus \epsilon_2\rangle , \qquad (14)$$

where $\epsilon_1, \epsilon_2 \in \{0,1\}$ and $\oplus$ is addition modulo 2. The (implicit) phase factor in the transformation is equal to 1. In this expression $\epsilon_1$ is the called the control bit and $\epsilon_2$ is the target bit. If $\epsilon_1 = 0$, the target bit remains unchanged; if $\epsilon_1 = 1$, the target bit flips. In the single-ion experiment of Ref. 53, the control bit is the quantized state of one mode of the ion's motion. If the motional state is $|n=0\rangle$, it is taken to be a $|\epsilon_1=0\rangle$ state; if the motional state is $|n=1\rangle$, it is taken to be a $|\epsilon_1=1\rangle$ state. The target states are two ground-hyperfine states of the ion, the $|\downarrow\rangle$ and $|\uparrow\rangle$ states of Sec. 5 with the identification here $|\downarrow\rangle \Leftrightarrow |\epsilon_2=0\rangle$ and $|\uparrow\rangle \Leftrightarrow |\epsilon_2=1\rangle$. Following the notation of Sec. 5, the CN operation is realized by applying three Raman laser pulses in succession:

(1a) A $\pi/2$ pulse is applied on the spin carrier transition. For a certain choice of initial phase, this corresponds to the operator $V^{\frac{1}{2}}(\pi/2)$ of Ref. 1.

(1b) A 2π pulse is applied on the first blue sideband transition between levels $|\uparrow\rangle$ and an auxiliary level $|aux\rangle$ in the ion (the $|F=2, M_F=0\rangle$ level in $^9Be^+$). This operator is analogous to the operator $U_n^{2,1}$ of Ref. 1. This operation provides the conditional dynamics for the controlled-not operation in that it changes the sign of the $|\uparrow\rangle|n=1\rangle$ component of the wavefunction but leaves the sign of the $|\uparrow\rangle|n=0\rangle$ component of the wavefunction unchanged.

(1c) A π/2 pulse is applied to the spin carrier transition with a 180° phase shift relative to step (1a). This corresponds to the operator $V^{\frac{1}{2}}(-\pi/2)$ of Ref. 1.

Steps (1a) and (1c) can be regarded as two resonant pulses (of opposite phase) in the Ramsey separated-field method of spectroscopy [58]. We can see that if step (b) is active (thereby changing the sign of the $|\uparrow\rangle|n=1\rangle$ component of the wave function) then a spin flip is produced by the (resonant) Ramsey fields. If step (1b) is inactive, the net effect of the Ramsey fields is to leave the spin state unchanged. This CN operation can be incorporated to provide an overall CN operation between two ions in an ensemble of N ions if we choose the ion oscillator mode to be a collective mode of the ensemble, such as the center-of-mass (COM) mode. Specifically, to realize a controlled-not $C_{m,k}$ between two ions (m = control bit, k = target bit), we first assume the COM is prepared in the zero-point state. The initial state of the system is therefore given by

$$\Psi = \left( \sum_{M_1=\downarrow,\uparrow} \sum_{M_2=\downarrow,\uparrow} \cdots \sum_{M_N=\downarrow,\uparrow} C_{M_1,M_2,\ldots M_N} |M_1\rangle_1 |M_2\rangle_2 \cdots |M_N\rangle_N \right) |0\rangle . \quad (15)$$

$C_{m,k}$ can be accomplished with the following steps:

(2a) Apply a π pulse on the red sideband of ion m (the assumption is that ions can be addressed separately [1]). This accomplishes the mapping $(\alpha|\downarrow\rangle_m + \beta|\uparrow\rangle_m)|0\rangle \rightarrow |\downarrow\rangle_m(\alpha|0\rangle - e^{i\phi}\beta|1\rangle)$, and corresponds to the operator $U_m^{1,0}$ of Ref. 1. In our experiments, we prepare states of the form $(\alpha|\downarrow\rangle + \beta|\uparrow\rangle)|0\rangle$ using the carrier transition (Sec. 5.1). We can then implement the mapping $(\alpha|\downarrow\rangle + \beta|\uparrow\rangle)|0\rangle \rightarrow |\downarrow\rangle_m(\alpha|0\rangle - e^{i\phi}\beta|1\rangle)$ by applying a red sideband π pulse. This is the "keyboard" operation for preparation of arbitrary motional input states for the CN gate of steps 1a-1c above. Analogous mapping of internal state superpositions to motional state superpositions was demonstrated in Ref. 53.

(2b) Apply the CN operation (steps 1a-1c above) between the COM motion and ion k.

(2c) Repeat step (2a).

Overall, $C_{m,k}$ provides the mappings $|\downarrow\rangle_m|\downarrow\rangle_k|0\rangle \rightarrow |\downarrow\rangle_m|\downarrow\rangle_k|0\rangle$, $|\downarrow\rangle_m|\uparrow\rangle_k|0\rangle \rightarrow |\downarrow\rangle_m|\uparrow\rangle_k|0\rangle$, $|\uparrow\rangle_m|\downarrow\rangle_k|0\rangle \rightarrow |\uparrow\rangle_m|\uparrow\rangle_k|0\rangle$, and $|\uparrow\rangle_m|\uparrow\rangle_k|0\rangle \rightarrow |\uparrow\rangle_m|\downarrow\rangle_k|0\rangle$ which is the desired logic of Eq. (14). Effectively, $C_{m,k}$ works by mapping the internal state of ion m onto the COM motion, performing a CN between the COM motion and ion n, and then mapping the COM state back onto ion m.

The resulting CN between ions m and k is not really different from the CN described by Cirac and Zoller, because the operations $V^{1/2}(\theta)$ and $U_m^{1,0}$ commute. A simplified scheme for quantum logic on ions, which does not require the auxiliary level, is discussed in Ref. 59.

*8.2 Quantum Registers and Schrödinger Cats*

The state during the CN operation is of the same form as that of Eq. (10). Both involve entangled superpositions and both are subject to the destructive effects of decoherence. Creation of Schrödinger cats like Eq. (10) is particularly relevant to the ion-based quantum computer because the primary source of decoherence will probably be due to decoherence of the $|n=0,1\rangle$ motional states during the logic operations.

*8.3    Perspective on ion quantum computation and quantum logic*

To be useful for factorization, a quantum computer must be able to factorize a 200 digit decimal number. This will require a few thousand ions and perhaps $10^9$ or more elementary operations [49]. Therefore, given the current state of the art (one ion and about 10 operations before decoherence), we should be skeptical. Decoherence will be most decisive in determining the fate of quantum computation. Decoherence from spontaneous emission will cause one fundamental limit to the number of operations possible [60,61]. The experiments can be expected to improve dramatically, but we must hope for more efficient algorithms or ways to patch them (such as error correction schemes [62-72]) before large scale factoring is possible. Possibilities and limitations with the ion trap scheme are being explored [60,73-78].
    Any quantum system that might be contemplated in quantum computation must be reproducible, stable, and well isolated from the environment. Quantum dots have the potential advantage of large-scale integration using microfabrication; however, at the present time, they suffer from lack of precise reproducibility and excessive decoherence. Trapped ions are reproducible and relatively immune to environmental perturbations - this is the reason they are candidates for advanced frequency standards [79]. In principle, higher information density and faster gate speeds (fundamentally limited by motional oscillation frequencies) could be achieved in ion traps by scaling down the size of the trap electrodes; however, we must then worry about excessive environmental coupling such as magnetic field perturbations caused by impurities and/or currents in the (nearby) trap electrodes. To avoid excessive coupling to the surrounding environment in any quantum logic scheme, the supporting matrix for quantum bits must therefore be reasonably separated from the qubits. In any scheme which uses atomic transitions, a scale size represented by present ion traps might be close to optimum.
    Factorization, discrete logs, sorting, and certain other mathematical computations appear to be the hardest problems that quantum logic might be applied to. One of the applications for quantum computation that Richard Feynman originally had in mind was to simulate quantum

mechanical calculations [80]. This idea is being explored again with new possibilities in mind [81-84]. Other potential applications might also be realized. At NIST, the original motivation for creation of entangled states was to fundamentally improve the signal-to-noise ratio in ion spectroscopy and frequency standards; we discuss this possibility in the next section.

## 9. QUANTUM LOGIC APPLIED TO SPECTROSCOPY

We conclude by discussing a possible application of quantum logic in the realm of atomic physics. This application has the advantage of being useful with a relatively small number of ions and logic operations.

Entangled atomic states can improve the quantum-limited signal-to-noise ratio in spectroscopy [20,85,86]. In spectroscopy experiments on N atoms, in which changes in atomic populations are detected, we can view the problem in the following way using the spin-½ analogy for two-level atoms. We assume that spectroscopy is performed by applying (classical) fields of frequency $\omega$ for a time $T_R$ according to the Ramsey method of separated fields [58]. After applying these fields, we measure the final state populations. For example, we might measure the operator $\tilde{N}_\downarrow$ corresponding to the number of atoms in the $|\downarrow\rangle$ state. In the spin-½ analogy, this is equivalent to measuring the operator $J_z$, since $\tilde{N}_\downarrow = J\tilde{I} - J_z$ where $\tilde{I}$ is the identity operator.

If all sources of technical noise are eliminated, the signal-to-noise ratio (for repeated measurements) is fundamentally limited by the quantum fluctuations in the number of atoms which are observed to be in the $|\downarrow\rangle$ state. These fluctuations can be called quantum "projection" noise [87]. If spectroscopy is performed on N initially uncorrelated atoms (for example, $\Psi(t=0) = \Pi_i |\downarrow\rangle_i$), the imprecision in a determination of the frequency of the transition is limited by projection noise to $(\Delta\omega)_{meas.} = 1/(NT_R\tau)^{½}$ where $\tau \gg T_R$ is the total averaging time. If the atoms can be initially prepared in entangled states, it is possible to achieve $(\Delta\omega)_{meas.} < 1/(NT_R\tau)^{½}$. Initial theoretical investigations [20,85] examined the use of correlated states which could achieve $(\Delta\omega)_{meas.} < 1/(NT_R\tau)^{½}$ when the population ($J_z$) was measured. More recent theoretical investigations [86] consider the initial state to be one where, after the first Ramsey pulse, the internal state is the maximally entangled state of Eq. (5). After applying the Ramsey fields, we measure the operator $\tilde{O} = \Pi_i \sigma_{zi}$ instead of $J_z$ (or $\tilde{N}_\downarrow$). For unit detection efficiency, we can achieve $(\Delta\omega)_{meas.} = 1/(N^2 T_R \tau)^{½}$ which is the maximum signal-to-noise ratio possible. The main reason for the reduced value of $(\Delta\omega)_{meas.}$ is that $\langle \tilde{O} \rangle = (-1)^N \cos[N(\omega - \omega_o)T_R]$; the derivative of this function with respect to applied frequency is N times larger than for the uncorrelated atom case. For an atomic clock where $T_R$ is fixed by other constraints, this means that the time required to reach a certain measurement precision (stability) is reduced by a factor of N relative to the uncorrelated-atom case. In terms of quantum computation, this amounts to a computation of the function $\cos(N(\omega - \omega_o)T)$. Of course, this computation has special significance for the measurement of $\omega_o$ (an intrinsic computer parameter) but otherwise is much better suited for a classical computer!

Cirac and Zoller [1] have outlined a scheme for producing the state in Eq.(5) using quantum logic gates. Using the notation of Sec. 5.1, we would first prepare the atoms in the state $\Psi(t=0) = \Pi_i |\downarrow\rangle_i |n=0\rangle$ and then apply a $\pi/2$ rotation to ion 1 to create the state $\Psi = 2^{-\frac{1}{2}}(|\downarrow\rangle_1 + e^{i\phi}|\uparrow\rangle_1)|\downarrow\rangle_2|\downarrow\rangle_3...|\downarrow\rangle_N|n=0\rangle$. We then apply the CN gate of Eq. (14) sequentially between ion 1 and ions 2 through N to achieve the state of Eq. (5). An alternative method for generating this state is described in Ref. 86.

## 10. ACKNOWLEDGMENTS

We gratefully acknowledge the support of the National Security Agency, the US Office of Naval Research, and the US Army Research Office. We thank E. Burt, J. Marquardt, D. Sullivan, and M. Young for helpful comments on the manuscript.

Figure Captions:

Fig. 1. The upper part of the figure shows a schematic diagram of the electrode configuration for a linear Paul-RF trap (rod spacing ≃ 1 mm). The lower part of the figure shows an image of a string of $^{199}$Hg$^+$ ions, illuminated with 194 nm radiation, taken with a UV-sensitive, photon-counting imaging tube. The spacing between adjacent ions is approximately 10 µm. The gaps in the string are occupied by impurity ions, most likely other isotopes of Hg$^+$, which do not fluoresce because the frequencies of their resonant transitions do not coincide with those of the 194 nm $^2S_{1/2}$ → $^2P_{1/2}$ transition of $^{199}$Hg$^+$.

Fig. 2. (a) Electronic (internal) and motional (external) energy levels (not to scale) of the trapped $^9$Be$^+$ ion, coupled by indicated laser beams R1 and R2. The difference frequency of the Raman beams R1 and R2 is set near $\omega_o/2\pi \simeq 1.250$ GHz, providing a two-photon coupling between the $^2S_{1/2}$(F=2, m$_F$=2) and $^2S_{1/2}$(F=1, m$_F$=1) hyperfine ground states (denoted by $|\downarrow\rangle$ and $|\uparrow\rangle$ respectively). The motional energy levels are depicted by a ladder of vibrational states separated in frequency by the trap frequency $\omega_x/2\pi \simeq 11$ MHz. The Raman beams are detuned $\Delta/2\pi \simeq -12$ GHz from the $^2P_{1/2}$(F=2, m$_F$=2) excited state. As shown, the Raman beams are tuned to the red sideband (see text). (b) Detection of the internal state is accomplished by illuminating the ion with $\sigma^+$-polarized "detection" beam D2, which drives the cycling $^2S_{1/2}$(F=2, m$_F$=2) → $^2P_{3/2}$(F=3, m$_F$=3) transition, and observing the scattered fluorescence. The vibrational structure is omitted from (b) since it is not resolved. Beam D1, also $\sigma^+$ polarized, provides spontaneous recycling from the $|\uparrow\rangle$ to $|\downarrow\rangle$ state.

Fig. 3. $P_\downarrow(\tau)$ for a coherent state driven by the first blue sideband interaction, showing collapse and revival behavior. The data are fitted to a coherent state distribution, yielding $\langle n \rangle = 3.1(1)$. The inset shows the results of inverting the time-domain data by employing a Fourier cosine transform at the known Rabi frequencies $\Omega_{n,n+1}$, fitted to a Poissonian distribution, yielding $\langle n \rangle = 2.9(1)$. Each data point represents an average of ≈4000 measurements, or 1 s of integration. (from Ref. 22)

Fig. 4. Surface and contour plot of the Wigner function of an approximate $|n=1\rangle$ number state.

Fig. 5. Reconstructed density matrix amplitudes of an approximate $1/\sqrt{2}$ ($|n=0\rangle$ - i $|n=2\rangle$) state. The amplitudes of the coherences indicate that the reconstructed density matrix is close to that of a pure state.